\documentclass{article}

\usepackage{arxiv}

\usepackage{multirow}
\usepackage{array}
\usepackage[ruled,vlined]{algorithm2e}

\usepackage{xcolor}

\usepackage[utf8]{inputenc} 
\usepackage[T1]{fontenc}    
\usepackage{hyperref} 
\usepackage{scalerel,stackengine}
\stackMath
\newcommand\reallywidehat[1]{%
\savestack{\tmpbox}{\stretchto{%
  \scaleto{%
    \scalerel*[\widthof{\ensuremath{#1}}]{\kern-.6pt\bigwedge\kern-.6pt}%
    {\rule[-\textheight/2]{1ex}{\textheight}}
  }{\textheight}%
}{0.5ex}}%
\stackon[1pt]{#1}{\tmpbox}%
}

\usepackage{graphicx}
\usepackage[export]{adjustbox}
\usepackage{xcolor}
\newcommand\Tstrut{\rule{0pt}{2.1ex}}       
\newcommand\Bstrut{\rule[-0.9ex]{0pt}{0pt}} 
\usepackage{subfigure}
\usepackage{colortbl}
\usepackage{paralist}
\usepackage{bbm}
\usepackage[arabic,farsi,english]{babel}
\usepackage{url}            
\usepackage{booktabs}       
\usepackage{amsfonts}       
\usepackage{nicefrac}       
\usepackage{microtype}      
\usepackage{lipsum}	
\usepackage{xcolor}
\title{For Women, Life, Freedom: A Participatory AI-Based Social Web Analysis of a Watershed Moment in Iran's Gender Struggles}

\author{
  Adel Khorramrouz\\
  {Rochester Institute of Technology}\\
  \texttt{ak8480@rit.edu} \\
\And
Sujan Dutta \\
  {Rochester Institute of Technology}\\
  \texttt{sd2516@rit.edu} \\
 \And
Ashiqur R. KhudaBukhsh\thanks{This work is accepted at IJCAI 2023 (AI for good track). Ashiqur R. KhudaBukhsh is the corresponding author.} \\
  {Rochester Institute of Technology} \\
  \texttt{axkvse@rit.edu}
} 

\begin{document}
\maketitle

\begin{abstract}

In this paper, we present a computational analysis of the Persian language Twitter discourse with the aim to estimate the shift in stance toward gender equality following the death of Mahsa Amini in police custody. We present an ensemble active learning pipeline to train a stance classifier. Our novelty lies in the involvement of Iranian women in an active role as annotators in building this AI system. Our annotators not only provide labels, but they also 
suggest valuable keywords for more meaningful corpus creation as well as provide short example documents for a guided sampling step. Our analyses indicate that Mahsa Amini's death triggered polarized Persian language discourse where both fractions of negative and positive tweets toward gender equality increased. The increase in positive tweets was slightly greater than the increase in negative tweets.  We also observe that with respect to account creation time, between the state-aligned Twitter accounts and pro-protest Twitter accounts,
pro-protest accounts are more similar to baseline Persian 
Twitter activity.

\end{abstract}


\section{Introduction}
\emph{Words are the only victors.}\\
-- Salman Rushdie; \emph{Victory City}; 2023.\\

On 16 September 2022, Mahsa Amini, a 22-year-old woman died under police custody in Iran. Reportedly, she was arrested because of not wearing her hijab (headscarf) properly. As media and police presented conflicting accounts of her death~\cite{MahsaAminiDeath}, Mahsa Amini's death enraged Persian (Farsi) Twitter users in an unprecedented manner~\cite{kermani2023mahsaamini}. The hashtag \FR{امینی}\_\FR{مهسا}\#  (\textit{\#MahsaAmini}) became one of the most repeated hashtags on Persian Twitter and initiated a Twitter protest where Iranians expressed their  grievances against the government like never before. Support and solidarity for gender equality poured over in from prominent world leaders~\cite{GuardianSupportBiden}, artists~\cite{AsgharFarhadi}, and sports personalities~\cite{IranChessPlayer} across the globe.

\#MahsaAmini was undoubtedly the overwhelming top-trending hashtag on Persian Twitter for months during the relentless protest. However, for a brief period of time, hashtags with an opposite stance toward the protest (e.g., \#ExecuteThem or \#ISupportKhamenei\footnote{Ali Khamenei is the second and current supreme leader of Iran who is in office since 1989.}) trended. Prior literature conjectured state-aligned trolling in Iran on Instagram~\cite{kargar2019state}. Also, social bot accounts' capability to spread extreme ideology is well-documented~\cite{stella2018bots,berger2015isis}.

Via a substantial corpus of 30.5 million tweets relevant to the protest, this paper makes three key observations:\\
1. \textit{The grievances of protesters against the current government mention a broad range of incidents spanning decades.}\\
2. \textit{With respect to account creation time, between the state-aligned Twitter accounts and pro-protest Twitter accounts, pro-protest accounts are more similar to baseline Persian Twitter activity.}\\
3. \textit{There was a noticeable shift in positive stance toward gender equality after the protests on Persian Twitter discourse.}\\

To our knowledge, no computational analysis relying on sophisticated natural language processing methods exists that has  examined gender equality in Persian social media discourse let alone at this unprecedented scale. That said, we believe our key contribution lies elsewhere. Our paper marks an important effort to include the stakeholders -- the Iranian women -- in this AI-building process. All examples in our supervised solution's training set are annotated by Iranian women. Our examples are thus grounded in cultural contexts and first-person experience about the gender struggles in Iran. 

Datasets addressing issues faced by vulnerable communities often end up being annotated by annotators with little or no documentation~\cite{guest-etal-2021-expert,ramesh-etal-2022-revisiting}. Since annotated examples often form the core of a supervised AI system, it is important to involve stakeholders in the annotation process. For example, Ramesh \textit{et al.}~\cite{ramesh-etal-2022-revisiting} present a lexicon of queer-related inappropriate words where one of the annotators identifies as queer. Similarly, Guest \textit{et al.}~\cite{guest-etal-2021-expert} present a misogyny dataset where the majority of the annotators identify as women.

Our annotators' role is not limited to mere annotation. Rather, they take an active role in guiding how to curate more meaningful data by suggesting suitable keywords to curate our dataset and providing a valuable seed set of examples to initiate an active learning pipeline. Our results indicate that the annotators' contributions yielded a richer seed set than a random baseline. 

At a philosophical level, we see this work as a part of the growing conversation of participatory AI~\cite{harrington2019deconstructing,delgado2022uncommon,bondi2021envisioning,birhane2022power}  where the goal is to develop systems for the people and by the people.  

\section{Datasets}\label{sec:Dataset}

As we already mention, \#MahsaAmini initiated a protest with global participation. Understandably, tweets in global languages such as English or French are likelier to reflect the global perspective on this issue. Given that Twitter is banned in Iran and users reportedly use VPNs to access Twitter~\cite{kermani2023mahsaamini}, considering geo-tagged tweets is not a reliable option either to understand and analyze the Persian perspective. Therefore, we restrict our analyses to only tweets authored in the Persian language. We assume that our choice of language can act as an effective filter to ensure our dataset is less likely to be diluted by the global discourse. We use Twitter's official language label as ground truth. 

We collect three corpora: $\mathcal{D}_\mathit{protest}$; $\mathcal{D}_\mathit{gender}$; and $\mathcal{D}_\mathit{baseline}$. 

Our dataset spans the time duration of Jan 15, 2022, to Jan 15, 2023\footnote{On January 7, 2023, two executions relevant to this protest happened~\cite{Jan2023Execution}. We thus set our end date one week after the executions.}.  We define the time period from January 15, 2022, to September 15, 2022, as $\mathcal{T}_\textit{before}$. We define the time period from September 16, 2022, to January 15, 2023, as $\mathcal{T}_\textit{after}$.  A short description follows next. Throughout the paper, if we use a Persian word or phrase, we present an English translation in parentheses following the word.  


\subsection{\colorbox{blue!25}{$\mathcal{D}_\mathit{gender}$}}
$\mathcal{D}_\mathit{gender}$ consists of  6,036,012 Tweets which has been posted by 700,189 unique users. 

\begin{compactenum}
    \item All tweets that have either ``\FR{زنان}" (\textit{women}) or ``\FR{دختر}'' (\textit{girl}).

    \item All tweets that have either ``\FR{ناموس}" which means \textit{the immediate female family members (daughter, mother, sister, wife) whom the male member of the family (father, brother, husband) should protect and sometimes control} or ``\FR{غیرت}" which means \textit{the positive form of jealousy that men have upon their female family members against other men}. These two search keywords were suggested by our annotators. 

    \item All tweets that have gender insult words against women ``{\FR{جنده}}'' and ``\FR{کصده}'' both indicating ``a prostitute'' or ``a promiscuous woman'' in a pejorative way (the second insult word mostly accompanies with \texttt{sister}).

    \item all tweets that have at least one word from the two following subsets: 
    \{``\FR{دختر}'' (\textit{girl}), ``\FR{زن}'' (\textit{woman}), ``\FR{خواهر}'' \textit{sister}\}; and \\ 
    \{``\FR {زندگی}" (\textit{life}), ``\FR{انقلاب}"(\textit{revolution}), ``\FR{حقوق}" (\textit{rights}), ``\FR{آزادی}" (\textit{freedom})\}.
\end{compactenum}

\subsection{\colorbox{blue!25}{$\mathcal{D}_\mathit{protest}$}}
$\mathcal{D}_\mathit{protest}$ consists of: 
\begin{compactenum}

    \item 

    tweets with \FR{امینی}\_\FR{مهسا}\# (\#\textit{MahsaAmini}) in them yielding 21,308,449 Tweets posted by 655,303 unique users.

    \item tweets that support government through the hashtag \FR{ای}\_\FR{خامنه}\_\FR{یا}\_\FR{لبیک}\#. This hashtag has been used 1,051,792 times by 71,484 unique users across the entire Twitter timeline accessible through the APIs.

    \item tweets that have the hashtag \FR{کنید}\_\FR{اعدام}\# which means \textit{execute them}. This hashtag has been used 11,292 times by 5,130 unique users across the entire Twitter timeline accessible through the APIs. 
\end{compactenum}

\subsection{\colorbox{blue!25}{$\mathcal{D}_\mathit{baseline}$}}
In order to estimate baseline Persian Twitter behavior, we consider five Persian stop words (\FR{به}, \FR{با}, \FR{در}, \FR{از}, \FR{که}) and collect 6,000 tweets per day (evenly distributed across the hours) that contain at least one of these stop words. Our dataset, $\mathcal{D}_\mathit{baseline}$, consists of 2,190,000 tweets. 

We compute the unigram distributions of subsets of $\mathcal{D}_\mathit{baseline}$ that was authored during $\mathcal{T}_\mathit{before}$ and $\mathcal{T}_\mathit{after}$. 
Table~\ref{tab:movers} lists the top 20 high-frequency non-stop words present when (1) we subtract the unigram distribution of $\mathcal{T}_\mathit{after}$ from the unigram distribution of $\mathcal{T}_\mathit{before}$ (left); and (2) we subtract the unigram distribution of $\mathcal{T}_\mathit{before}$ from the unigram distribution of $\mathcal{T}_\mathit{after}$ (left). In plain English, these are the words that appeared more frequently during one period and much less frequently during the other. From the right column of Table~\ref{tab:movers}, we note that several of these words are not indicative of civic unrest while the left column does not indicate similar unrest. We conduct a similar experiment to track shift in high-frequency hashtag usage between the two time periods. We again observe that even in the baseline Persian Twitter discourse, $\mathcal{T}_\mathit{after}$  
showed several hashtags relevant to the protest. 
\begin{table}[htb]
{
\small
\begin{center}
     \begin{tabular}{| p{7.5cm}  | p{7.5cm} |}
    \hline
    \Tstrut More presence during $\mathcal{T}_\mathit{before}$ & More presence during $\mathcal{T}_\mathit{after}$  \\
     \hline \Tstrut

``\FR{خوش}" (\textit{happy}),
``\FR{انسان}" (\textit{human}),
``\FR{زبان}" (\textit{language}),
``\FR{آقا}" (\textit{M.R}),
``\FR{نمیدونم}" (\textit{i do not know}),
``\FR{توییتر}" (\textit{twitter}),
``\FR{پسر}" (\textit{boy}),
``\FR{قبول}" (\textit{ok}),
``\FR{گوش}" (\textit{ear}),
``\FR{حد}" (\textit{limit}),
``\FR{خواب}" (\textit{sleep}),
``\FR{جدید}" (\textit{new}),
weareoneEXO,
``\FR{دلیل}"(\textit{reason}),
``\FR{عشق}"(\textit{love}),
``\FR{روسیه}"(\textit{Russia}),
``\FR{الله}"(\textit{god(Allah)}),
``\FR{نظرم}"(\textit{my opinion}),
``\FR{حس}"(\textit{feeling}),
``\FR{خوبی}"(\textit{goodness}),

\Bstrut&      

``\FR{مادر}" (\textit{mother})
``\FR{کشته}" (\textit{killed})
``\FR{صدای}" (\textit{voice})
``\FR{مرگ}" (\textit{death})
``\FR{آزادی}" (\textit{freedom})
``\FR{خون}" (\textit{blood})
``\FR{جمهوری}" (\textit{republic})
``\FR{خبر}" (\textit{news})
``\FR{اعدام}" (\textit{execution})
``\FR{بخاطر}" (\textit{for sake of})
``\FR{هشتگ}" (\textit{hashtag})
``\FR{ادامه}" (\textit{continue})
``\FR{خانواده}" (\textit{family})
``\FR{نظام}" (\textit{can be translated to regime but not exact})
``\FR{شهر}" (\textit{city})
``\FR{لطفا}" (\textit{please})
``\FR{انقلاب}" (\textit{revolution})
``\FR{اسم}" (\textit{name})
``\FR{ج.ا}" (\textit{I.R (stands for Islamic republic which represents regime)})
``\FR{خیابون}" (\textit{street})
\\
    \hline
    \end{tabular}
\end{center}
\caption{\small{Biggest shift in token usage in $\mathcal{D}_\mathit{baseline}$ between $\mathcal{T}_\mathit{before}$ and $\mathcal{T}_\mathit{after}$}.}
\label{tab:movers}}
\end{table}


\begin{table}[htb]

{
\small
\begin{center}
     \begin{tabular}{| p{7.5cm}  | p{7.5cm} |}
    \hline
    \Tstrut Top ten hashtags during $\mathcal{T}_\mathit{before}$ & Top ten hashtags during $\mathcal{T}_\mathit{after}$ \\
     \hline \Tstrut
EXO,
``\FR{مکتب\_امید}" (\textit{the\_hope\_attitude}),
``\FR{ما\_ملت\_امام\_حسینیم}" (\textit{we\_are\_nation\_of\_Imam\_Hossein}),
``\FR{ماه\_امید}" (\textit{month\_of\_hope}),
``\FR{ایران\_قوی}" (\textit{strong\_Iran}),
``\FR{حب\_الحسین\_یجمعنا}" (\textit{love\_of\_Hossein\_gathers\_us}),
``\FR{اوکراین}" (\textit{Ukraine}),
EXO (\textit{in Korean}),
``\FR{اللهم\_عجل\_لوليك\_الفرج}" (\textit{Oh God, please fasten merging relief (Imam Zaman)for us}),
``\FR{عید\_امید}" (\textit{Hope\_Eyd}),

\Bstrut&

``\FR{مهسا\_امینی}" (\textit{mahsa\_amini}),
``\FR{اعتصابات\_سراسری}" (\textit{Nationwide\_strikes}),
OpIran,
MahsaAmini,
StopHazaraGenocide,
IRGCterrorists,
``\FR{نیکا\_شاکرمی}" (\textit{Nika\_Shakarami}),
``\FR{زن\_زندگی\_آزادی}" (\textit{women\_life\_freedom}),
Mahsa\_Amini,
``\FR{اعتراضات\_سراسری}" (\textit{Nationwide\_protests}),
``\FR{محسن\_شکاری}" (\textit{Mohsen\_Shekari}),

 \\
    \hline
    \end{tabular}
\end{center}
\caption{\small{Shift in top hashtags present in $\mathcal{D}_\mathit{baseline}$ between $\mathcal{T}_\mathit{before}$ and $\mathcal{T}_\mathit{after}$.}}
\label{tab:moversHashtag}}
\end{table}

\begin{table}[htb]
{\small
\begin{center}
     \begin{tabular}{|p{5cm} |p{5cm} | p{3cm}  |}
    \hline
  Line from Baraye & Translation & Percentage of match \\
    \hline
    \FR{برای دختری که آرزو داشت پسر بود} & {for a girl who wished she were a boy } & 24.86\\
     \hline 
    \FR{برای آزادی} & {for freedom} & 15.72\\
     \hline 
    \FR{برای خواهرم خواهرت خواهرامون} & {for my sister, your sister, our sisters} & 9.76\\
     \hline 
    \FR{برای این همه شعار های تو خالی} & {For all these meaningless slogans} & 7.91\\
     \hline 
    \FR{برای این بهشت اجاری} & {For this forced "heaven"} & 7.16\\
     \hline 
    \FR{تغییر مغز ها که پوسیدن	برای} & {for changing rusted minds} & 7.09\\
     \hline 
    \FR{برای چهره ای که میخنده} & {For smiling faces} & 4.33\\
     \hline 
    \FR{برای کودک زباله گرد و آرزو هاش} & {for child labor and their crushed dreams} & 3.82\\
     \hline 
     
    \end{tabular}
    
\end{center}
\caption{\small{Percentage of \textit{because of} tweets that matched with individual lines in Grammy-winning song Baraye by Shervin Hajipour.}}
\label{tab:Baraye}}
\vspace{-0.2cm}
\end{table}

\section{Baraye -- Because Of}\label{sec:BecauseOf}

A large fraction of tweets of $\mathcal{D}_\textit{protest}$  contains a phrase (\textit{because of}). These tweets were an outlet for Persian Twitter users to vent their frustrations about the situation. Specifically, the tweets aimed at answering a reason for the protests. The tweets expressing a complex collection of emotions on why the current government has failed the nation captured the imagination of Shervin Hajipour, a talented Iranian singer who won the 2023 Grammy award for his song Baraye (\textit{because of}). Each line of this song starts with \texttt{because of} and paints a picture of Persian hope and despair. We collect 1.92 million tweets from $\mathcal{D}_\textit{protest}$ with the phrase \textit{because of}. We train a FastText~\cite{bojanowski2017enriching} word embedding on $\mathcal{D}_\textit{protest}$ and for each tweet, we compute the line in the Baraye song that is the nearest neighbor in the embedding space. Table~\ref{tab:Baraye} lists the top 10 lines from the Baraye song that matched with the tweets. 

While the poignant song by Hajipour brilliantly captures Persian aspirations and struggles, Table~\ref{tab:BarayeTopTrigrams} indicates there is much more to Persian angst than what the song could hold. The second most common trigram indicates the access barrier to Twitter. Most prominent social media platforms are blocked in Iran~\cite{sohrabi2021new} and reportedly, Iranians primarily take recourse to VPNs to participate in the social web~\cite{kermani2023mahsaamini}. We observe that Navid Afkari\footnote{\url{https://www.hrw.org/news/2020/09/12/iran-suddenly-executes-wrestler-navid-afkari}}, an executed Iranian wrestler, is mentioned among the top trigrams. Multiple structural failures got also mentioned in the common trigrams~\cite{BuildingCollapse}. From the Cinema Rex fire incident that happened in 1978~\cite{ali2018reacting}, to the attacks on student-dormitory in 1999~\cite{BuildingCollapse}, to the current reality of web censorship,  the most striking takeaway from Table~\ref{tab:BarayeTopTrigrams} is perhaps the time duration between the events people mentioned.  

\begin{table}[htb]
{\small
\begin{center}
     \begin{tabular}{|l |l |}
    \hline
  Trigram & Translation \\
    \hline

    \FR{حدیث نجفی مهسا} & Hadis Najafi Mahsa [Amini]\\
     \hline
    \FR{زن زندگی آزادی} & women life freedom\\
     \hline
     
    \FR{دسترسی توییتر ندارد} & have not Twitter access\\
     \hline 
    \FR{رمزی میشود آزادی} & [your name] will become [the] symbol [of] freedom \\
     \hline 
    \FR{نامت رمزی میشود} & your name will become symbol\\
     \hline 
    \FR{بختیاری نوید افکاری} & [Pouya] Bakhtiari Navid Afkari\\
     \hline 
    \FR{های ریخته شده} & spilled [blood]\\
     \hline 
    \FR{پلاسکو سانچی کولبران} & Plasco Sanchi kolbar  \\
    \hline 
    \FR{متروپل آبان پویا} &  Metropoll Aban (November) Pouya [Bakhtiari] \\
     \hline 
    \FR{سینما رکس کوی } &  Cinema Rex kouye [university]\\
   \hline

    \end{tabular}
    
\end{center}
\caption{\small{Top ten trigrams from \textit{because of} tweets. We omit two spam trigrams (e.g., please follow me) aimed at gaining followers. Hadis Najafi was killed by a gunshot during the Mahsa Amini protest.}}
\label{tab:BarayeTopTrigrams}}
\end{table}

\section{Account Creation Time}

\begin{figure*}[htb]
\centering
\includegraphics[trim={0 0 0 0},clip, width=0.90\textwidth]{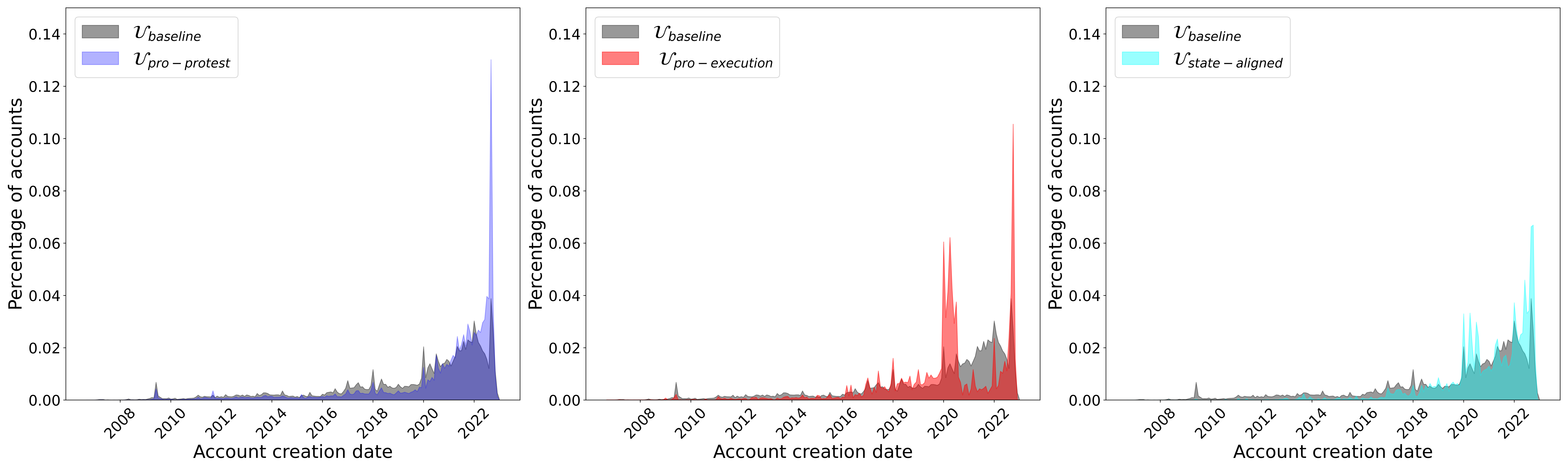}
\caption{\small{Distributions of account creation dates of different user sets.}} 

\label{fig:AccountCreationDate}
\end{figure*}

Prior literature has examined state-aligned trolling in Iran on Instagram platforms~\cite{kargar2019state}. In this section, we present an analysis based on account creation time. We define four sets of users: $\mathcal{U}_\textit{pro-protest}$; $\mathcal{U}_\textit{state-aligned}$; $\mathcal{U}_\textit{pro-execution}$; and $\mathcal{U}_\textit{baseline}$. $\mathcal{U}_\textit{pro-protest}$ represents all users who used the hashtag \#MahsaAmini (indicating support for Mahsa Amini) at least once in our dataset. $\mathcal{U}_\textit{state-aligned}$ represents all users who used the hashtag \#ISupportKhamenei (indicating support for Ali Khamenei) at least once in our dataset. $\mathcal{U}_\textit{pro-execution}$ represents the set of users who used the hashtag \#ExecuteThem at least once in our dataset. Finally, $\mathcal{U}_\textit{baseline}$ is the set of unique users who contributed to $\mathcal{D}_\textit{baseline}$. We compute the account creation time for each user at the granularity of months and obtain normalized histograms for each of these sets.  Figure~\ref{fig:AccountCreationDate} illustrates the account creation temporal distributions of the four user sets. All subsets exhibit a sharp spike around September 2022, however $\mathcal{U}_\textit{pro-execution}$ exhibits two different spikes. In fact, in September 2022, Google Trends indicates one of the most popular search queries from Iran was ``\FR{دانلود توییتر}'' (\textit{download Twitter}).  

Table~\ref{tab:KLDivergence} computes the KL divergence of the account creation time distributions with respect to $\mathcal{U}_\textit{baseline}$. We observe that the distribution of $\mathcal{U}_\textit{pro-protest}$ is closest to $\mathcal{U}_\textit{baseline}$ while $\mathcal{U}_\textit{baseline}$ is the farthest. The order remains unchanged with other distributional distance measures (e.g., Bhattacharyya distance). Our qualitative findings remain unchanged even if we limit $\mathcal{U}_\textit{state-aligned}$ and $\mathcal{U}_\textit{pro-execution}$ to only those user accounts that used these hashtags during $\mathcal{T}_\textit{before}$ and $\mathcal{T}_\textit{after}$.   

\begin{table}[htb]
{\small
\begin{center}
     \begin{tabular}{|l |c |}
    \hline
  User Set & KL-Divergence \\
    \hline
  $\mathcal{U}_\textit{pro-protest}$ &  0.15
     \\ 
     \hline 
   $\mathcal{U}_\textit{state-aligned}$ &  0.25
     \\
     \hline
     $\mathcal{U}_\textit{pro-execution}$ &  0.46
     \\
     \hline
    \end{tabular}
    
\end{center}
\caption{\small{KL-divergence of the distribution of account creation time for different user subsets with respect to the distribution of account creation time of $\mathcal{U}_\mathit{baseline}$.}}
\label{tab:KLDivergence}}
\end{table}

\section{Annotation}

All our annotation was conducted by four different annotators. All annotators identify as Iranian women and are fluent speakers of Persian. All of them have undergraduate degrees.

\paragraph{Annotation task.} Our text prediction task is to predict the stance toward gender equality. For each tweet, we ask the annotator: \emph{does this short document indicate a positive, neutral, or negative stance toward gender equality?}

\paragraph{Inter-rater Agreement.} Between any two annotators, we have at least 500 overlapping samples. Across all rounds of annotation, the Cohen's $\kappa$ ranged from  0.41 to 0.52. On a misogyny annotation task, Guest \textit{et al.}~\cite{guest-etal-2021-expert} reported  Fleiss’ $\kappa$ of
0.48 and the Krippendorf’s alpha as 0.49. Sanguinetti \textit{et al.}~\cite{sanguinetti2018italian} report category-wise $\kappa=0.37$
for offence and $\kappa=0.54$ for hate. We further note that our observed inter-rater agreement is higher than Gomez \textit{et al.}~\cite{gomez2020exploring} ($\kappa$ = 0.15)
and Fortuna and Nunes~\cite{fortuna2018survey} ($\kappa$ = 0.17). 

\paragraph{Disagreement resolution.} Since our task is likely to be subjective, resolving disagreements has to be grounded in the literature. Prior literature has considered diverse approaches to resolving inter-annotator disagreements (e.g., majority voting~\cite{davidson2017automated,wiegand2019detection} or third objective instance~\cite{DBLP:conf/ranlp/GaoH17}). We resolve any disagreement in the following manner. For positives and neutrals, we only consider consensus labels. Following~Golbeck \textit{et al.}~\cite{golbeck2017large}, if any annotator marks an example as negative and the other annotator marks it as negative or neutral, we consider the aggregate label as negative. In order to ensure the anonymity of the annotators, we do not conduct any post-annotation adjudication step to resolve disagreements. 

\subsection{Toward Participatory AI}

A notable feature in our work is the active involvement of Iranian annotators in both corpus creation and annotation. Our annotators helped us in the following two ways. 

\paragraph{Search keywords.} At a deeper level, which data could contain relevant information may require a clear understanding of the social realities. While constructing $\mathcal{D}_\mathit{gender}$, choosing \texttt{woman} or \texttt{girl} and gendered insults as search keywords required little cultural context. However, our annotators suggested nuanced keywords such as ``\FR{ناموس}'' and  ``\FR{غیرت}'' to be included in our list of search keywords. Recall that, ``\FR{ناموس}'' means the immediate female family members (e.g., daughter, mother, sister, or wife) whom the male members (e.g., father, brother, or husband) should protect and sometimes control; and ``\FR{غیرت}" means a positive form of jealousy that men have upon their female family members against other men.   

\paragraph{Seed set.} A notable feature of our work is the active involvement of Iranian women in the annotation process, where they not only provide labels but also present important representative short documents to construct meaningful seed sets during the guided sampling step described in Section~\ref{sec:ActiveLearning}.


\section{Active Learning Pipeline}~\label{sec:ActiveLearning}

\noindent\textbf{Research question}: \emph{Is there a noticeable change in support for gender equality in Persian Twitter discourse before and after the demise of Mahsa Amini while in police custody?}

To estimate the support for gender equality in Persian Twitter discourse, we build a robust classifier detecting content supportive of gender equality. Since hashtag hijacking~\cite{hadgu2013political} is a common phenomenon where users with opposite views may use the most-popular hashtag to express an opposite stance, our goal is to predict the stance toward gender equality from tweet texts only.  

We first estimate to which extent tweets supporting gender equality are present in $\mathcal{D}_\mathit{baseline}$. We randomly sample 500 tweets weighing both $\mathcal{T}_\mathit{before}$ and $\mathcal{T}_\mathit{after}$ equally (i.e., 250 from each time slice). In addition, we randomly sample 1,000 tweets from $\mathcal{D}_\mathit{gender}$ weighing equally $\mathcal{T}_\mathit{before}$ and $\mathcal{T}_\mathit{after}$. Table~\ref{tab:labelDistribution} summarizes the label distribution. We note that a large fraction of $\mathcal{D}_\mathit{baseline}$ consists of neutral tweets. 

In order to construct a dataset that is diverse and representative of the unlabeled pool, we present an active learning pipeline that consists of well-known sampling steps. A short description of active learning follows next.

\begin{table}
\centering
  \begin{tabular}{lcccccc}
    \toprule
    \multirow{2}{*}{Dataset} &
      \multicolumn{3}{c}{$\mathcal{T}_\mathit{before}$} &
      \multicolumn{3}{c}{$\mathcal{T}_\mathit{after}$}  \\
      & Positive &  Neutral & Negative & Positive &  Neutral & Negative \\
      \midrule
    
    $\mathcal{D}_\mathit{gender}$ & 11.9\% & 51.9\% & 36.1\% & 30.4\% & 35.3\% & 34.2\% \\
    $\mathcal{D}_\mathit{baseline}$ & 0.4\% & 98.3\% & 1.2\% & 2.8\% & 93.4\% & 3.6\% \\
    \bottomrule
  \end{tabular}
  \caption{\small{Label distribution of the first stage of annotation (random sampling) during the seed set construction.}}
  \label{tab:labelDistribution}
\end{table}

\subsection{Background}
\emph{Active Learning} is a powerful and well-established form of supervised machine learning technique~\cite{settles2009active}. It is characterized by the interaction between the learner, aka the classifier, and the teacher (oracle or labeler or annotator) during the learning process. At each iteration, the learner employs a sampling strategy to select an unlabeled sample (unlabeled samples) and requests the supervisor to label it (them) in agreement with the target concept. The data set is augmented with the newly acquired label, and the classifier is retrained on the augmented data set. The sequential label-requesting and re-training process continues until some halting condition is reached (e.g., annotation budget is expended or the classifier has reached some target performance). At this point, the algorithm outputs a classifier, and the objective for this classifier is to closely approximate the (unknown) target concept in the future. The key goal of active learning is to reach a strong performance at the cost of fewer labels. Since retraining the model and running inference on a large, unlabeled pool is computationally costly, prior literature has examined the trade-offs present in a batch active learning setting~\cite{yang2013buy}. In this work, we follow the batch active learning setting.


\subsection{Seed Set Construction}
\paragraph{Random Sampling.} In order to capture a diverse set of examples, we randomly select 1,000 samples from $\mathcal{D}_\mathit{gender}$ and 500 samples from $\mathcal{D}_\mathit{baseline}$. Table~\ref{tab:labelDistribution} indicates that solely relying on $\mathcal{D}_\mathit{baseline}$ to construct the seed set will result in extreme class imbalance with very few positives and negatives and predominantly neutrals. Sampling from $\mathcal{D}_\mathit{gender}$ might yield slightly more positives (and negatives), however, a keyword-based starting point runs the risk of biasing the whole active learning pipeline. In what follows, we present a guided sampling approach similar to Palakodety~\textit{et al.}~\cite{palakodety2020voice}.

\paragraph{Guided Sampling.} When faced with the challenge to find high-quality positive examples championing the Rohingya community, Palakodety~\textit{et al.}~\cite{palakodety2020voice} proposed a document-embedding-based, guided sampling method where annotators provide example short documents conforming to a given label. We employ a similar technique where we asked three annotators to provide five examples each indicating positive and negative stances toward gender equality. For each example, we select 25 unique nearest neighbors in the document embedding space from the unlabeled pool giving equal weightage to tweets from $\mathcal{T}_\mathit{before}$ and $\mathcal{T}_\mathit{after}$. This yields 750 samples. Upon annotation and resolving disagreements, we obtain 166 positives, 145 negatives, and 231 neutrals. We note that our sampling method yielded substantially more positives (and negatives) than the random sampling baseline.

\begin{table*}[htb]
{
\small
\begin{center}
     \begin{tabular}{|p{0.40\textwidth}|p{0.40\textwidth}|}
     \hline
     Seed examples produced by annotators & Translation \\
    \hline
\cellcolor{blue!15}\FR{چرا یک زن که آدم بالغی است و خودش عقل داره و توانایی اینو داره که بره یه کشور دیگه، باید از یک نفر مرد دیگه اجازه خروج بگیره؟ حالا با هر نسبتی
} &  \cellcolor{blue!15}\cellcolor{blue!15}\textit{Why does a woman, who is an adult, and has the ability to go to another country, need to seek another man's permission to leave regardless of whatever the relation is?
} \\
\hline
 \cellcolor{blue!15}\FR{باید این همه تفاوت در استانداردهای مان را کنار بیاندازیم و اگر رفتاری را برای مردها مناسب می دانیم بر زنها هم روا بداریم و درک کنیم که هیچ فرقی نیست بین مرد و زنی که توی یک مهمانی سیگار می کشند یا می رقصند یا با صدای بلند می خندند
} & \cellcolor{blue!15}\emph{We have to put aside all the differences in our standards and
if  we consider  a behavior appropriate for men, we should allow women as well, and  understand that there is no difference between a man and a woman who smoke or dance or laugh loud at a party.}\\
\hline
\cellcolor{blue!15}\FR{خانمها باید هنگام طلاق حقوق برابر با آقایون داشته باشند یعنی حق حضانت بچه و حق طلاق و... را داشته باشند.} & \cellcolor{blue!15}\emph{When it comes to divorce, women should have the same rights as men, that means, they should have the right to custody of the child and the right to divorce, etc.
}\\
    \hline

\cellcolor{red!15}\FR{آدم این وضعیت پوشش این دخترای بی حجاب رو که توی خیابون میبینه بیشتر مطمئن میشه که اینا آزادی رو توی همون لخت شدن میبینن و غیر از این دنبال هیچی نیستن.} & \cellcolor{red!15}\emph{When you see these hijabless girls' attire in the street, you can be sure that they see freedom in being naked and they are not looking for anything else.} \\
\hline
 \cellcolor{red!15}\FR{دخترا اگه خیلی فوتبال دوست دارن خب بشینن تو خونه ببینن، نمیخواد برای جلب توجه برن استادیوم
} & \cellcolor{red!15}\emph{If girls like football so much, they can sit at home and watch it. They don't have to go to the stadium and attract attention.
.}\\
\hline
\cellcolor{red!15}\FR{معلومه که باید دیه زن نصف مرد باشه؛ مردا نون آور خونواده هستن و تازه دیه به زن و بچهه میرسه به نفعشونم هست، گیر الکی به قانون ندید
} & \cellcolor{red!15} \emph{It is obvious that a woman's blood money should be half a man's. Men are the breadwinners of the family, and the blood money goes to the women and children. Do not tinker around with the law for no good reason.}\\
    \hline
    \end{tabular}
    
\end{center}
\caption{\small{Random sample of seed examples presented by annotators. Blue indicates a positive stance toward gender equality and red indicates a negative stance toward gender equality.}}
\label{tab:SeedExamples}}
\end{table*}

\begin{table*}[htb]
{
\small
\begin{center}
     \begin{tabular}{|p{0.41\textwidth}|p{0.41\textwidth}|}
     \hline
    Examples obtained through guided sampling & Translation \\
    \hline
\cellcolor{blue!15}\FR{حق حضانت‌ برای تو!
 درد زایمان برای من 
نام خانوادگی برای تو! 
زحمت خانواده برای من 
سند خانه به نام تو!
 بیگاری خانه برای من 
چهار عقد برای تو!
 حسرت عشق برای من 
هزار صیغه برای تو!
 حکم سنگسار برای من 
هوس برای تو!
 عفاف برای من این 
بود برابری حقوق زن و مرد؟} & \cellcolor{blue!15}\emph{Custody right for you!
	  Labor pain for me
	Surname for you!
	Family trouble for me
	The document of the house in your name!
	  Hardwork in home for me
	Four marriages for you!
	  Missing love for me
	A thousand concubines for you!
	  A sentence of stoning for me
	Lust for you!
	 chastity for me
	Are these equal rights for men and women?}  \\
\hline
 \cellcolor{blue!15}\FR{حجاب خانم های خونی ، نسبی، ثبتی من نه تنها به کسی مربوط نیست بلکه به بنده هم مربوط نیست زن یک انسان مستقل هست دست از مالکیت بردارید زنان برده هیچ مردی نیستند آنان که خود را تسلیم مرد میکنند نه تنها به خود بلکه به زنان دیگر هم خیانت میکنند
} &\cellcolor{blue!15}\emph{
Neither me nor anyone can tell my relatives if they should wear a hijab or not.
Women are not slaves of men. Those women who submit themselves to men are not only betraying themselves, but also betraying other women.}\\
\hline
\cellcolor{blue!15}\FR{میدونی دیگه ایران زن نمیتونه جدا شه مگر دلایلی بیاره که ….! درحالی که یک مرد بای دیفالت اون حق و داره . همسرشما داشته از اول. و شما تازه به دستش اوردی و هم سطح شدی.\ldots} & \cellcolor{blue!15}\emph{You know, Iranian women can't ask for divorce unless they give reasons.! While a man by default has that right. Your husband  had it from the beginning. And you just earned it and  leveled up.\ldots
}\\
    \hline

\cellcolor{red!15}\FR{مردا بهتر از زنان کار می کنند
} & \cellcolor{red!15}\emph{Men work better than women.} \\
\hline
 \cellcolor{red!15}\FR{خود زنها عرضه ندارن حقشون رو بگیرن ، مردها مقصرند ؟!! چند بار تا حالا شنیدی که زنها برای گرفتن حق حضور در استادیوم برن جلوی فدراسیون تجمع کنن ؟ ولی از مردها انتظار دارن که به استادیوم نرن تا از زنها حمایت بشه . متاسفانه اکثریت جامعه زنان ایران فقط غر زدن رو بلدن .} & \cellcolor{red!15}\emph{Is this men's fault that women themselves do not have ability to take their rights?!! How many times have you heard that women gather in front of the federation to get the right to attend the stadium? But they expect men not to go to the stadium to support women. Unfortunately, the majority of Iranian women only know how to nag
.}\\
\hline
\cellcolor{red!15}\FR{اون آزادی بیشتری که بعضی ها می‌خوان شخصیت زن رو منحصر به زیبایی های ظاهری می‌کنه و این هم برای خودش و هم برای جامعه ضرر داره} & \cellcolor{red!15}\emph{The more freedom people want for women, the more women's character gets overshadowed by physical beauty, which is harmful to both herself and the society.}\\
    \hline
    \end{tabular}
    
\end{center}
\caption{\small{Random sample of tweet texts retrieved through guided sampling. Blue indicates a positive stance toward gender equality and red indicates a negative stance toward gender equality.}}
\label{tab:GuidedSampling}}
\end{table*}

Table~\ref{tab:SeedExamples} presents a few randomly selected positive and negative seed examples provided by our annotators. We observe that the examples are grounded in women's cultural struggle in Iran~\cite{kazemzadeh2002islamic}. Beyond discussions around hijab, inequality in marital and inheritance law~\cite{doherty2021divorce}, restrictions on activities such as visiting stadiums to watch football~\cite{GuardianBlueGirl,abtahi2022bluegirl} echoed in these examples.   

Table~\ref{tab:GuidedSampling} lists a random sample of retrieved tweet texts when we used the guided sampling method. This table shows that not only we found more positives (and negatives) than the random baseline, but the tweet texts also exhibit richness, diversity, and nuance.

Overall, we obtain 343 positives, 440 negatives, and 1,051 neutrals from the random sampling and guided sampling step. In what follows, we describe two well-known sampling strategies that we employ to further expand our dataset.

\subsection{Certainty and Uncertainty Sampling}

\paragraph{Certainty sampling.}
Since our goal is to use the trained model for a social inference task, it is important to rectify high-confidence misclassifications. Minority class certainty sampling has found its use in rectifying high-confidence misclassifications involving short documents such as movie reviews and messages~\cite{sindhwani2009uncertainty,attenberg2010unified}; search queries~\cite{khudabukhsh2015building}; and comments on YouTube videos~\cite{palakodety2020voice}. We conduct certainty sampling for the positive class and select 750 instances that the model predicts as positive with the highest confidence. We also conduct certainty sampling for the negative class and select 750 instances that the model predicts as negative with the highest confidence. In this step, we obtain 338 positives, 345 negative, and 487 neutrals.

\paragraph{Uncertainty sampling.} Uncertainty sampling is one of the most well-known sampling strategies used in active learning~\cite{settles2009active}. Since we have multiple label categories in our prediction task, we use margin sampling, an active learning variant designed for multiple labels~\cite{scheffer2001active}. In this step, we sample 1,500 examples. Upon annotation and resolving the disagreements, we obtain 115 positives, 247 negatives, and 819 neutrals.  

\begin{table}[htb]
{\small
\begin{center}
     \begin{tabular}{|l |l | c  |}
    \hline
  Data & Model & Macro F$_1$ \\
    \hline 
 $\mathcal{D}_\textit{seed}$  &  $\mathcal{M}_\textit{seed}$ & 67.09 $\pm$ 1.46 \\
    \hline 
$\mathcal{D}_\textit{seed}$ $\cup$ $\mathcal{D}_\textit{certainty}$  &   $\mathcal{M}_\textit{certainty}$ & 69.28 $\pm$ 0.61 \\ 
    \hline
$\mathcal{D}_\textit{seed}$ $\cup$ $\mathcal{D}_\textit{certainty}$ $\cup$  $\mathcal{D}_\textit{uncertainty}$&      $\mathcal{M}_\textit{uncertainty}$ & 73.27 $\pm$ 1.87 \\ 
     \hline 
    \end{tabular}
    
\end{center}
\caption{\small{Performance comparison of  models trained on various stages of our active learning pipeline. $\mathcal{M}_\textit{seed}$  denotes a popular Persian language model \cite{farahani2021parsbert} trained on $\mathcal{D}_\textit{seed}$. Subsequent models are fine-tuned on top of this. For all models trained in this paper, performance is reported over five different training runs on a fixed evaluation set of randomly sampled 400 instances from our annotated dataset ensuring no overlap between train and tests.}}
\label{tab:Classification}
}
\end{table}

\begin{figure}[htb]
\centering
\includegraphics[trim={0 0 0 0},clip, width=0.65\textwidth]{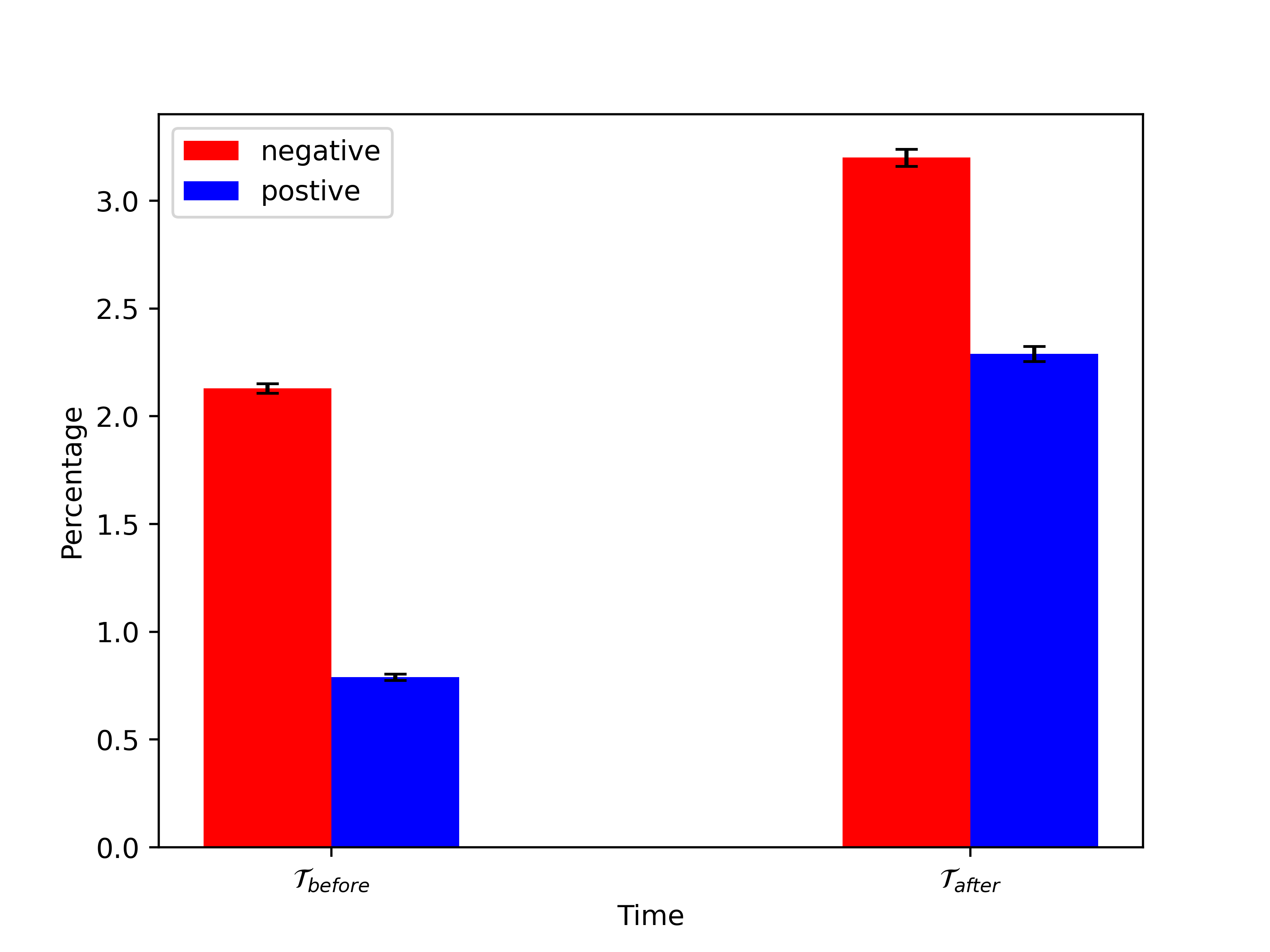}
\caption{{Temporal trend of tweets expressing positive and negative stance toward gender inequality on $\mathcal{D}_\mathit{baseline}$.}} 

\label{fig:Inference}
\end{figure}

To summarize, our active learning pipeline consists of the following steps:
\begin{compactenum}
\item Construct an initial seed set by randomly sampling from  $\mathcal{D}_\textit{random}$, and $\mathcal{D}_\textit{gender}$, and using guided sampling ($\mathcal{D}_\textit{seed}:$ 343 positives, 440 negatives, and 1,051 neutral instances) using random sampling.  
\item Conduct certainty sampling on the positive class and certainty sampling on the negative class ($\mathcal{D}_\textit{certainty}: $ 338 positives, 345 negatives, and 487 neutral instances). 
\item Finally, conduct uncertainty sampling (margin sampling) ($\mathcal{D}_\textit{uncertainty}: $ 115 positive, 247 negative, and 819 neutral instances). 
\end{compactenum}

Overall, we obtain 796 positive, 1,032 negative, and 2,357 neutral examples. 

\subsection{Model Performance and Analysis}

Table~\ref{tab:Classification} summarizes the performance of our trained models. The performance improves at each active learning step and we finally achieve a Macro F$_1$ performance of 73.27\%. We note that if we train a binary classifier with just the positive class and the negatives and neutrals clubbed together as the \textit{notPositive} class, it is possible to achieve slightly better performance (Macro F$_1$: 77.76 $\pm$ 2.29).

To track shifts in stance toward gender equality, we run inference using $\mathcal{M_\textit{certainty}}$ on $\mathcal{D}_\textit{baseline}$. Figure~\ref{fig:Inference} indicates that the discourse became more polarized during $\mathcal{T}_\mathit{after}$ with both percentages of tweets expressing positive and negative stances increasing. However, we also observe that the increase in positive discourse (by a factor of 2.89) is greater than the increase in negative discourse (by a factor of 1.50).




\section{Discussions}

In this paper, we present the first-ever computational analysis (to the best of our knowledge) of the stance toward gender equality in Persian Twitter discourse 
following a watershed moment in Iran's history. Our analyses reveal that the grievances of Persian Twitter users against the government span decades and the protest following Mahsa Amini's death perhaps presented an outlet for the angst harbored for a long time. Second, we observe that the distribution of account creation time can present important signals. We find that with respect to account creation time, pro-execution and state-aligned user sets are distributionally different from baseline Persian Twitter users.

We follow an ensemble active learning pipeline to construct a robust classifier that detects stance toward gender equality. As a step towards participatory AI, our annotators take an active role in building our machine learning model. There is a growing concern that our ML conversations barely include marginalized community which can further widen the gap of AI-haves and AI-have-nots. All our annotators are Iranian women, with first-person experience of gender struggles. Their role in our system was far more profound than typical annotators. In a guided sampling step, they provided seed examples to expand our dataset lending cultural grounding. They also suggest important keywords to curate our dataset.   


Table~\ref{tab:BarayeTopTrigrams} suggests that a major Iranian grievance is limited access to the internet. Free and fair access to Twitter was many users' wish. While we were working on this paper, Twitter as a platform underwent several significant changes. With the deprecation of academic Twitter and developer accounts being monetized, at this point, it is unclear how much of our collected data would be accessible in the future and at what cost. This is a curious juxtaposition of a community longing for access to a platform to voice their concerns while the very same platform is limiting academic researchers' access to study global politics. 

On top of the current uncertainties surrounding Twitter, the inherently transient nature of the social web, censorship, and fear of persecution can contribute to missing content for post-hoc analyses. In that sense, our paper is a humble attempt to preserve a vulnerable chunk of the social web that chronicled a watershed moment in gender struggle in Persian history.

\section*{Ethical Statement}
We use publicly available tweets collected using academic Twitter API. Since our data is highly sensitive, we only conduct aggregate analyses without revealing personally identifiable information. We also do not conduct any post-annotation adjudication steps that are typical to many annotation tasks to ensure the privacy of the annotators. 

We trained our model on top of a large language model. 
Several lines of recent research have indicated that large language models have a wide range of biases that reflect the texts on which they were originally trained, and which may percolate to downstream tasks~\cite{bender2021dangers}.



\end{document}